\documentclass[conference]{IEEEtran}
\IEEEoverridecommandlockouts

\usepackage{cite}
\usepackage{tabularx}
\usepackage{dsfont}
\usepackage{amsmath,amssymb,amsfonts}
\usepackage{algorithmic}
\usepackage{algorithm}
\usepackage{bm}
\usepackage{graphicx}
\usepackage[caption=false,font=footnotesize,labelfont=rm,textfont=rm]{subfig}
\usepackage{textcomp}
\usepackage{xcolor}
\def\BibTeX{{\rm B\kern-.05em{\sc i\kern-.025em b}\kern-.08em
    T\kern-.1667em\lower.7ex\hbox{E}\kern-.125emX}}
\begin{document}

\title{Energy-Efficient RF Chain Activation for Uplink mmWave Cell-Free Massive MIMO Systems: A Minimum-Cost Maximum-Flow Approach
}

\author{\IEEEauthorblockN{Bo Yu, Abdol Hamid Aghvami, Mohammad Shikh-Bahaei}
\IEEEauthorblockA{Department of Engineering, King's College London, London WC2R 2LS, United Kingdom \\
\{bo.1.yu, hamid.aghvami, m.sbahaei\}@kcl.ac.uk}
}

\maketitle

\begin{abstract}
With the incessant demand for green and sustainable communications, energy efficiency (EE) has become a cornerstone of sixth-generation (6G) network design, especially for cell-free millimeter-wave (mmWave) massive multiple-input multiple-output (MIMO) systems, which can achieve significant throughput gains by leveraging large-scale antenna arrays. Adaptive radio frequency (RF) chain activation (ARFA) is a key technique for reducing power consumption without severely degrading the achievable rate, thereby enhancing the total EE of 6G MIMO systems. In this paper, we propose a minimum-cost maximum-flow (MCMF)-based ARFA scheme to reduce unnecessary power consumption and maximize the total EE of uplink mmWave cell-free massive MIMO systems. Simulation results indicate that the proposed approach can significantly improve the total EE while maintaining a relatively high achievable rate and acceptable computational complexity.
\end{abstract}


\section{Introduction}
Millimeter wave (mmWave) has been deployed in fifth-generation (5G) and 5G-Advanced (5G-A) mobile networks, and continues to play a critical role in sixth-generation (6G) systems. The prominent advantage of mmWave is the abundance of spectrum resources, enabling significantly higher channel capacity than the sub-6 GHz bands. However, mmWave propagation suffers from severe path loss due to high frequencies, which constrains its commercial deployment regarding cellular coverage and power consumption~\cite{6834753}.

Massive multiple-input multiple-output (MIMO) provides an effective solution to compensate for the large path loss in mmWave channels through the enhanced beamforming capabilities of large-scale antenna arrays equipped at base stations (BSs)~\cite{11037786}. Furthermore, the authors in~\cite{7827017} pioneer the concept of cell-free massive MIMO, where numerous distributed access points (APs) connected to a central processing unit (CPU) simultaneously serve a much smaller number of user equipment (UEs) over the same time-frequency resource based on channel state information (CSI), achieving significantly improved throughput compared with conventional cellular massive MIMO. While cellular massive MIMO employs point-to-point transmission between a UE and its assigned BS, cell-free massive MIMO leverages the cooperation of multiple APs to serve each UE, thereby eliminating cell boundaries and ensuring uniform per-user quality of service (QoS). However, deploying such massive MIMO systems---whether cellular or cell-free---poses major challenges regarding energy consumption, thermal loads, and hardware costs attributable to the extensive use of large-scale antenna arrays and their associated radio frequency (RF) front-end (RFFE) components, such as power amplifiers (PAs) and low-noise amplifiers (LNAs).

As energy efficiency (EE) has emerged as a critical metric in wireless network design, extensive research on system EE improvement has been conducted~\cite{8424051},\cite{9497709},\cite{8097026},\cite{9780003},\cite{al2024energy},\cite{9849036},\cite{11192810},\cite{10747281}. Given the architectural differences, EE optimization approaches, such as BS sleep control and RF chain activation in~\cite{8424051}, and resource allocation in~\cite{9497709}, are primarily tailored for traditional cellular massive MIMO, whereas dedicated schemes for cell-free massive MIMO have been investigated in~\cite{8097026},\cite{9780003},\cite{al2024energy},\cite{9849036},\cite{11192810},\cite{10747281}. Specifically, in~\cite{8097026}, the authors propose power allocation and AP selection schemes, optimizing the downlink EE in sub-6 GHz. In \cite{9780003}, the authors propose adaptive RF chain activation (ARFA) schemes in the mmWave uplink, which dynamically activate/deactivate RF chains and their cascaded components, i.e., analog-to-digital converters (ADCs) and phase shifters (PSs), under imperfect CSI. In~\cite{al2024energy}, a matching method based on Hungarian and genetic algorithms has been applied in mmWave uplink systems to selectively activate RF chains. Additionally, machine learning (ML)-based algorithms have been devised to solve the EE optimization problem~\cite{9849036}, \cite{11192810}, \cite{10747281}. However, \cite{9849036} optimizes the system EE in the sub-6 GHz uplink,~\cite{11192810} aims for the sub-6 GHz downlink, and~\cite{10747281} considers the mmWave downlink. To the best of our knowledge, existing studies on EE optimization for cell-free massive MIMO predominantly focus on AP and/or antenna selection and power allocation in sub-6 GHz downlink systems, with limited attention paid to RF chain activation in mmWave uplink scenarios.

In this work, we propose a joint AP and RF chain activation approach by solving a minimum-cost maximum-flow (MCMF) problem to adaptively activate APs and assign an optimal set of RF chains to maximize the total EE of the uplink mmWave cell-free massive MIMO system. The specific contributions of this paper can be summarized as follows:
\begin{itemize}
\item We investigate a practical scenario considering imperfect channel conditions, such as estimation errors and the sparse multipath effect in mmWave propagation.
\item We propose an MCMF-based ARFA scheme to solve the formulated EE maximization problem.
\item The computational complexity of the proposed approach is compared with existing schemes.
\end{itemize}

The remainder of this article is organized in the following manner. Section~\ref{sec:system_arch_and_model} describes the system architecture and model. Section~\ref{sec:problem_and_solution} introduces the problem formulation, proposed solution, and complexity analysis. Section~\ref{sec:perf_eval} presents the simulation setup and performance evaluation. Finally, Section~\ref{sec:conclusion} concludes this work.

\section{System Architecture and Model}
\label{sec:system_arch_and_model}

\subsection{System Model}
\label{subsec:system_model}
\begin{figure}[htbp]
	\centering
	\includegraphics[width=3.5in]{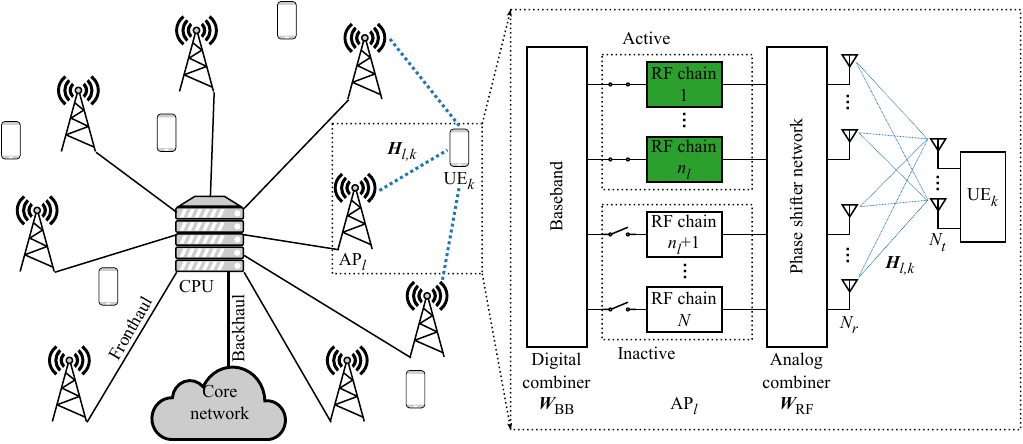}
	\caption{Cell-free massive MIMO system architecture with hybrid analog-digital beamforming and adaptive RF chain activation.}
	\label{cell_free_massive_mimo}
\end{figure}

As shown in Fig.~\ref{cell_free_massive_mimo}, we consider an uplink mmWave cell-free massive MIMO system with hybrid beamforming, where $L$ APs simultaneously and jointly serve $K$ UEs. Each AP is equipped with $N_r$ receive antennas and $N$ RF chains, where $N \le N_r$, while each UE comprises $N_t$ transmit antennas. All APs are connected to a CPU via fronthaul links to constitute a radio access network (RAN), which is then connected to a core network via backhaul links. In addition, a fully-connected phase shifter (PS) network is employed at each AP for analog beamforming, wherein each of the $N$ RF chains is connected to all $N_r$ antennas through $N N_r$ PSs. 

Furthermore, the system is assumed to operate in time division duplex (TDD) mode within low-speed mobility environments. Under TDD operation, the propagation channels are quasi-static over a coherence interval of $\tau_c$ time-frequency samples, where $\tau_c = T_c B_c$ is the product of the coherence time $T_c$ and the coherence bandwidth $B_c$. Each coherence interval $\tau_c$ is divided into three phases: $\tau_p$ for uplink pilot transmission, $\tau_d$ for downlink  data transmission, and $\tau_u$ for uplink data transmission. Note that the fronthaul links are assumed to be error-free and offer sufficiently high capacity to support perfect transmission between the APs and the CPU.

The received signal vector $\boldsymbol{y}_l \in \mathbb{C}^{KN_t \times 1}$ at the $l$-th AP, obtained by processing the transmitted signals from all $K$ UEs through their respective channel matrices $\boldsymbol{H}_{l,k} \in \mathbb{C}^{N_r \times N_t}$, followed by an analog combining matrix $\boldsymbol{W}_{\text{RF},l} \in \mathbb{C}^{N_r \times N}$ and a digital combining matrix $\boldsymbol{W}_{\text{BB},l} \in \mathbb{C}^{N \times KN_t}$, can be modeled as~\cite{9780003}
\begin{align}
	\label{eq_system1}
	\boldsymbol{y}_l &= \sqrt{\rho}\sum_{k=1}^K\boldsymbol{W}_{\text{BB},l}^H\boldsymbol{W}_{\text{RF},l}^H\boldsymbol{H}_{l,k}\boldsymbol{x}_{k}+\boldsymbol{W}_{\text{BB},l}^H\boldsymbol{W}_{\text{RF},l}^H\boldsymbol{z}_l \notag \\ 
	&= \sqrt{\rho}\sum_{k=1}^K\boldsymbol{W}_{\text{BB},l}^H\boldsymbol{W}_{\text{RF},l}^H\hat{\boldsymbol{H}}_{l,k}\boldsymbol{x}_{k}+\boldsymbol{W}_{\text{BB},l}^H\boldsymbol{W}_{\text{RF},l}^H\hat{\boldsymbol{z}}_l,
\end{align}
where ${\boldsymbol{x}_{k} \in \mathbb{C}^{N_t \times 1}}$ denotes the symbol vector transmitted by the $k$-th UE with ${\mathbb{E}[\|\boldsymbol{x}_k\|^2]} = 1$, and $\rho$ represents the average transmit power. The term $\boldsymbol{z}_l \sim \mathcal{CN}(\boldsymbol{0}, \sigma_n^2 \boldsymbol{I}_{N_r})$ is the independent and identically distributed (i.i.d.) additive white Gaussian noise (AWGN) vector, consisting of random variables (RVs) with zero mean and variance $\sigma_n^2 = -174\ \text{dBm/Hz} + 10\log_{10}(B) + \mathrm{NF}$, where $B$, $\mathrm{NF}$, and $\boldsymbol{I}_{N_r}$ are the system bandwidth, noise figure, and identity matrix, respectively. Furthermore, according to the channel estimation error model, we have $\boldsymbol{H}_{l,k} = \hat{\boldsymbol{H}}_{l,k} + \boldsymbol{\Delta}_{l,k}$, where the elements of the estimation error matrix $\boldsymbol{\Delta}_{l,k} \in \mathbb{C}^{N_r \times N_t}$ are i.i.d. Gaussian RVs with zero mean and variance $\varepsilon_{l,k}^2$, and it is uncorrelated with $\hat{\boldsymbol{H}}_{l,k} \in \mathbb{C}^{N_r \times N_t}$ based on the minimum mean square error (MMSE) estimation~\cite{8097026}. Notably, $\hat{\boldsymbol{H}}_{l,k} \in \mathbb{C}^{N_r \times N_t}$ and $\hat{\boldsymbol{z}}_l=\sqrt{\rho}\sum_{k=1}^K \boldsymbol{\Delta}_{l,k}\boldsymbol{x}_{k}+\boldsymbol{z}_l \sim \mathcal{CN}(\boldsymbol{0}, (\sigma_n^2 + \rho N_t \varepsilon^2)\boldsymbol{I}_{N_r})$ are the corresponding channel and noise estimation, where we assume an identical estimation error variance across all APs, given by $\varepsilon^2 = {\textstyle \sum_{k=1}^K} \varepsilon_{l,k}^2, \forall l$.

Finally, the locally detected signals $\boldsymbol{y}_l, \forall l$ are sent to the CPU via a fronthaul link for further processing. The composite received signal available at the CPU can be written as~\cite{9780003}
\begin{align}
	\label{eq_system2}
	\boldsymbol{y} &= \sqrt{\rho}\boldsymbol{W}_{\text{BB}}^H\boldsymbol{W}_{\text{RF}}^H\boldsymbol{H}\boldsymbol{x}+\boldsymbol{W}_{\text{BB}}^H\boldsymbol{W}_{\text{RF}}^H\boldsymbol{z} \notag \\ 
	&= \sqrt{\rho}\boldsymbol{W}_{\text{BB}}^H\boldsymbol{W}_{\text{RF}}^H\hat{\boldsymbol{H}}\boldsymbol{x}+\boldsymbol{W}_{\text{BB}}^H\boldsymbol{W}_{\text{RF}}^H\hat{\boldsymbol{z}},
\end{align}
where we define $\boldsymbol{y} = [\boldsymbol{y}_1^T,\dots,\boldsymbol{y}_L^T]^T \in \mathbb{C}^{LKN_t \times 1}$, $\boldsymbol{x} = [\boldsymbol{x}_1^T,\dots,\boldsymbol{x}_K^T]^T \in \mathbb{C}^{KN_t \times 1}$,  $\boldsymbol{z} = [\boldsymbol{z}_1^T,\dots,\boldsymbol{z}_L^T]^T \in \mathbb{C}^{LN_r \times 1}$, $\hat{\boldsymbol{z}} = [\hat{\boldsymbol{z}}_1^T,\dots,\hat{\boldsymbol{z}}_L^T]^T \in \mathbb{C}^{LN_r \times 1}$, $\boldsymbol{H} = [\boldsymbol{H}_1^T,\dots,\boldsymbol{H}_L^T]^T \in \mathbb{C}^{LN_r \times KN_t}$, $\hat{\boldsymbol{H}} = [\hat{\boldsymbol{H}}_1^T,\dots,\hat{\boldsymbol{H}}_L^T]^T \in \mathbb{C}^{LN_r \times KN_t}$, $\boldsymbol{W}_{\text{RF}}=\text{diag}(\boldsymbol{W}_{\text{RF},1},\dots,\boldsymbol{W}_{\text{RF},L}) \in \mathbb{C}^{LN_r \times LN}$, and $\boldsymbol{W}_{\text{BB}}=\text{diag}(\boldsymbol{W}_{\text{BB},1},\dots,\boldsymbol{W}_{\text{BB},L}) \in \mathbb{C}^{LN \times LKN_t}$. Besides, we can derive $\hat{\boldsymbol{z}} = \sqrt{\rho}\boldsymbol{\Delta x} + \boldsymbol{z} \sim \mathcal{CN} (\boldsymbol{0}, (\sigma_n^2 + \rho N_t\varepsilon^2)\boldsymbol{I}_{LN_r})$, by defining $\boldsymbol{\Delta} = [\boldsymbol{\Delta}_1,\dots,\boldsymbol{\Delta}_L]^T \in \mathbb{C}^{LN_r \times KN_t}$.

\subsection{Channel Model}
\label{subsec:channel_model}
The cluster-based geometric Saleh-Valenzuela (S-V) channel model is widely adopted in mmWave cellular systems~\cite{9780003},\cite{al2024energy},\cite{6717211},\cite{6847111}. For our considered cell-free massive MIMO system, the channel matrix $\boldsymbol{H}_{l,k} \in \mathbb{C}^{N_r \times N_t}$ between the $l$-th AP and the $k$-th UE is expressed as~\cite{9780003},\cite{6847111}
\begin{equation}
	\label{eq_channel}
	{\boldsymbol{H}}_{l,k} = \sqrt{\frac{G_r G_t N_r N_t}{\beta_{l,k}P_{l,k}}}
	\sum_{p=1}^{P_{l,k}}\alpha_{l,k}^{(p)}{\boldsymbol{a}}_{r}(\phi_{l,k}^{(p)}){\boldsymbol{a}}_{t}^H(\psi_{l,k}^{(p)}),
\end{equation}
where $G_r$ and $G_t$ denote the antenna element gains at the $l$-th AP and the $k$-th UE, respectively. The term $\beta_{l,k}$ represents the large-scale fading coefficient between the $l$-th AP and the $k$-th UE, which is given in dB as~\cite{7109864}
\begin{equation}
	\label{eq_pathloss}
	\beta_{l,k}[\text{dB}] = 10\log_{10}\left(\frac{4\pi{d_0}}{\lambda}\right)^2+10n_\text{pl}\log_{10}\left(\frac{d_{l,k}}{d_0}\right)+\chi_{l,k},
\end{equation}
with the close-in free-space reference distance $d_0=1$~m, the average path loss exponent $n_\text{pl}$, and the shadow fading component $\chi_{l,k} \sim \mathcal{N}(0, \sigma_{\text{sf}}^2)$. The variable $P_{l,k}$ is the number of effective propagation paths corresponding to a limited number of mmWave scatterers between the $l$-th AP and the $k$-th UE, and the small-scale fading coefficient of the $p$-th path is given by $\alpha_{l,k}^{(p)} \sim \mathcal{CN}(0, 1)$. The vectors $\boldsymbol{a}_r(\phi_{l,k}^{(p)}) \in \mathbb{C}^{N_r \times 1}$ and $\boldsymbol{a}_t(\psi_{l,k}^{(p)}) \in \mathbb{C}^{N_t \times 1}$ represent the normalized receive and transmit antenna array response vectors, corresponding to the azimuth angle of arrival (AoA) and angle of departure (AoD), respectively. Finally, we assume a uniform linear array (ULA) configuration for all AP and UE antennas, such that the antenna array response vector $\boldsymbol{a}(\phi)$ can be expressed as $\boldsymbol{a}(\phi) = \frac{1}{\sqrt{N_a}}[1,e^{j\frac{2\pi}{\lambda}d_s\sin(\phi)}, \dots, e^{j(N_a-1)\frac{2\pi}{\lambda}d_s\sin(\phi)}]^T$, where $N_a$, $\lambda$, and $d_s$ are the number of antennas, carrier wavelength, and antenna spacing, respectively~\cite{6847111}.

\subsection{Achievable Rate}
\label{subsec:achievable_rate}
Since $\hat{\boldsymbol{H}}$ and $\boldsymbol{\Delta}$ are independent, the hybrid combiners $\boldsymbol{W}_{\text{RF}}$ and $\boldsymbol{W}_{\text{BB}}$ are also independent of $\boldsymbol{\Delta}$. As indicated in \eqref{eq_system2}, the channel estimate $\hat{\boldsymbol{H}}$ is treated as the actual channel $\boldsymbol{H}$, and the term $\boldsymbol{W}_{\text{BB}}^H\boldsymbol{W}_{\text{RF}}^H\hat{\boldsymbol{z}}$ is considered the effective noise.

To reduce computational overhead and eliminate global CSI exchange, we adopt a decentralized hybrid beamforming (D-HBF) architecture~\cite{7389996}. Let $\mathcal{A} \subseteq \{1, \dots, L\}$ be the set of active APs with cardinality $|\mathcal{A}|$. Then, for the $l$-th AP ($l \in \mathcal{A}$) with $n_l$ activated RF chains ($1 \le n_l \le N$), the optimal analog combiner $\boldsymbol{W}_{\text{RF},l}^\star = [\boldsymbol{w}_{\text{RF},l,1}^\star, \dots, \boldsymbol{w}_{\text{RF},l,n_l}^\star] \in \mathbb{C}^{N_r \times n_l}$ is constructed locally by performing singular value decomposition (SVD) on the estimated channel $\hat{\boldsymbol{H}}_l \in \mathbb{C}^{N_r \times KN_t}$ such that $\hat{\boldsymbol{H}}_l = \boldsymbol{U}_l \boldsymbol{\Lambda}_l \boldsymbol{V}_l^H$. Furthermore, the optimal analog weight vector $\boldsymbol{w}_{\text{RF},l,n}^\star$ corresponding to the $n$-th active RF chain is extracted by quantizing the phase of the $n$-th dominant left singular vector $[\boldsymbol{U}_l]_{:,n}$ as $w_{\text{RF},l,i,n}^\star = \frac{1}{\sqrt{N_r}} e^{j \arg( [\boldsymbol{U}_l]_{i,n})}, \forall i \in \{1,\dots,N_r\}, n \in \{1,\dots,n_l\}$ under the constant modulus constraint $|w_{\text{RF},l,i,n}^\star| = \frac{1}{\sqrt{N_r}}$, where $[\boldsymbol{U}_l]_{i,n}$ represents the $(i,n)$-th element of the matrix $\boldsymbol{U}_l$.

After local analog combining, the effective channel block at the $l$-th AP is derived as $\boldsymbol{H}_{\text{eff},l} = \boldsymbol{W}_{\text{RF},l}^{\star H} \hat{\boldsymbol{H}}_l \in \mathbb{C}^{n_l \times KN_t}$. By stacking the effective channel blocks from all active APs, the global composite effective channel matrix available at the CPU is formed as $\boldsymbol{H}_{\text{eff}} = [\boldsymbol{H}_{\text{eff},l}^T]_{l \in \mathcal{A}}^T \in \mathbb{C}^{(\sum_{l \in \mathcal{A}} n_l) \times KN_t}$.

Consequently, under the centralized processing framework at the CPU, the Gram matrix of the global effective channel is constructed to evaluate the total achievable rate per unit bandwidth, i.e., the spectral efficiency (SE) in bit/s/Hz, which is expressed as
\begin{equation}
	\label{eq_rate1}
	R_\text{sum} = \log_2 \det \left( \boldsymbol{I}_{KN_t} + \gamma \boldsymbol{H}_{\text{eff}}^H \boldsymbol{H}_{\text{eff}} \right),
\end{equation}
where $\gamma = \frac{\rho}{\sigma_n^2 + \rho N_t\varepsilon^2}$. Furthermore, for local digital processing at the $l$-th AP, the optimal digital combiner $\boldsymbol{W}_{\text{BB},l}^\star \in \mathbb{C}^{n_l \times KN_t}$ can be derived from the MMSE solution as~\cite{7389996}
\begin{equation}
	\label{eq_rate2}
	\boldsymbol{W}_{\text{BB},l}^\star = \boldsymbol{J}_l^{-1}\boldsymbol{W}_{\text{RF},l}^{\star H}\hat{\boldsymbol{H}}_l^H,
\end{equation}
where $\boldsymbol{J}_l = \boldsymbol{W}_{\text{RF},l}^{\star H}\hat{\boldsymbol{H}}_l\hat{\boldsymbol{H}}_l^H\boldsymbol{W}_{\text{RF},l}^\star + \frac{1}{\gamma}\boldsymbol{W}_{\text{RF},l}^{\star H}\boldsymbol{W}_{\text{RF},l}^\star$.

\subsection{Power Consumption}
\label{subsec:power_consumption}
The total power consumption of the considered cell-free massive MIMO system can be modeled as~\cite{8097026},\cite{9780003}
\begin{equation}
\label{eq_power1}
P_{\text{total}} = \sum_{k=1}^K \left(P_{\text{TX},k} + P_{\text{CP},k}\right) + \sum_{l=1}^L \left(P_{\text{fix},l} + P_{\text{BF},l} + P_{\text{FH},l}\right),
\end{equation}
where $P_{\text{TX},k}$ and $P_{\text{CP},k}$ represent the transmit power and additional power consumed by other circuit components at the $k$-th UE, respectively. The terms $P_{\text{fix},l}$, $P_{\text{BF},l}$, and $P_{\text{FH},l}$ correspond to the fixed, beamforming, and fronthaul power consumption at the $l$-th AP. Specifically, $P_{\text{TX},k}$ is given as $P_{\text{TX},k} = \rho\sigma_n^2 \frac{\mathbb{E}[\|\boldsymbol{x}_k\|^2]}{\eta_k}=\frac{\rho\sigma_n^2}{\eta_k}$, where $\eta_k \in (0,1]$ represents the PA efficiency of the $k$-th UE. Regarding the hybrid beamforming architecture, each antenna is connected to an LNA and two mixers, and each RF chain requires an ADC, $N_r$ PSs, and a switch (SW), as illustrated in Fig.~\ref{cell_free_massive_mimo}. Thus, $P_{\text{BF},l}$, which linearly depends on the $N_r$ receive antennas and $n_l$ active RF chains, can be expressed as $P_{\text{BF},l} = N_r p_{\text{BF},1} + n_l p_{\text{BF},2}$, where $p_{\text{BF},1} = p_{\text{LNA}} + 2p_{\text{MIX}}$ and $p_{\text{BF},2} = N_r p_{\text{PS}} + p_{\text{RF}} + p_{\text{ADC}} + p_{\text{SW}}$ denote the power consumption of the antenna-related and RF chain-related components, respectively. According to~\cite{8097026}, we have $P_{\text{FH},l} = P_{\text{FH,max}}\frac{R_{\text{FH},l}}{C_{\text{FH},l}}$, where $P_{\text{FH,max}}$ represents the maximum power required for the fronthaul link traffic at the full capacity $C_{\text{FH},l}$, and $R_{\text{FH},l}$ denotes the actual fronthaul link rate between the $l$-th AP and the CPU given by $R_{\text{FH},l} = \frac{2KN_t\tau_u\alpha_l}{T_c}$, where $\alpha_l$ is the number of quantization bits at the $l$-th AP.

For simplicity, assuming that all UEs have identical power parameters, we set $\eta_k = \eta$ and $P_{\text{CP},k} = P_{\text{CP}}$. Likewise, we assume $P_{\text{fix},l} = P_{\text{fix}}$, $\alpha_l = \alpha$, $C_{\text{FH},l} = C_{\text{FH}}$, $P_{\text{FH},l} = P_{\text{FH}}$, and $R_{\text{FH},l} = R_{\text{FH}}$ for all APs. Note that the $l$-th AP consumes $P_\text{fix}$ even when inactive, while $P_\text{FH}$ and $P_{\text{BF},l}$ are consumed only when it is activated. As a result, the total power consumption in \eqref{eq_power1} can be rewritten as
\begin{equation}
\label{eq_power2}
P_{\text{total}} = P_0 + \left|\mathcal{A}\right| P_{\text{FH}} + \left|\mathcal{A}\right| N_r p_{\text{BF},1} + p_{\text{BF},2} \sum_{l \in \mathcal{A}} n_l,
\end{equation}
where $P_0 = \frac{K \rho \sigma_n^2}{\eta} + K P_{\text{CP}} + L P_{\text{fix}}$ represents the constant power consumption term.

\subsection{Energy Efficiency}
\label{subsec:energy_efficiency}
Based on the above overall achievable rate and system power consumption models, the total EE (in bit/J) of the uplink mmWave cell-free massive MIMO system is defined as the ratio of the total throughput to the total power consumption, expressed as
\begin{equation}
	\label{eq_ee}
	\mathrm{EE} = \frac{B R_\text{sum}}{P_\text{total}},
\end{equation}
where $B$ is the system bandwidth.

\section{Problem Formulation and Solution}
\label{sec:problem_and_solution}
\subsection{Problem Formulation}
\label{subsec:prob_form}
The resource assignment problem to maximize the total system EE can be formulated as
\begin{align}
	\label{eq_prob1}
	\max_{\{a_l\}, \{n_l\}, \{w_{\text{RF},l,i,n}\}} & \mathrm{EE} = \frac{B R_{\text{sum}}\left(\{a_l\}, \{n_l\}, \{w_{\text{RF},l,i,n}\}\right)}{P_{\text{total}}\left(\{a_l\}, \{n_l\}\right)} \notag \\ 
	\text{s.t.} \quad & a_l \in \{0, 1\}, \quad \forall l, \notag \\ 
	& a_l \leq n_l \leq a_l N, \quad \forall l, \notag \\ 
	& |w_{\text{RF},l,i,n}| = \frac{1}{\sqrt{N_r}}, \quad \forall l, i, n,
\end{align}
where the optimization variables consist of the AP activation indicators $\{a_l\}$, the corresponding numbers of active RF chains $\{n_l\}$, and the analog combining weights $\{w_{\text{RF},l,i,n}\}$, with $l \in \{1,\dots,L\}$, $i \in \{1,\dots,N_r\}$, and $n \in \{1,\dots,n_l\}$ indexing the APs, receive antennas, and active RF chains, respectively.

\subsection{Proposed Solution}
\label{subsec:prop_sol}
To solve the well-known mixed-integer non-linear programming (MINLP) problem in \eqref{eq_prob1}, we employ Dinkelbach's method to transform the fractional objective function into an equivalent subtractive form~\cite{dinkelbach1967nonlinear}. Let $q$ be the auxiliary variable representing the system EE. Consequently, for a given $q$, the parameterized problem can be reformulated as
\begin{align}
	\label{eq_prob2}
	\max_{\{a_l\}, \{n_l\}, \{w_{\text{RF},l,i,n}\}} & F(q) = B R_{\text{sum}}\left(\{a_l\}, \{n_l\}, \{w_{\text{RF},l,i,n}\}\right) \notag \\ 
	& \phantom{F(q) = } - q P_{\text{total}}\left(\{a_l\}, \{n_l\}\right) \notag \\ 
	\text{s.t.} \quad & a_l \in \{0, 1\}, \quad \forall l, \notag \\ 
	& a_l \leq n_l \leq a_l N, \quad \forall l, \notag \\ 
	& |w_{\text{RF},l,i,n}| = \frac{1}{\sqrt{N_r}}, \quad \forall l, i, n.
\end{align}
The overall solution follows an iterative framework as summarized in Algorithm~\ref{alg:dinkelbach_sv_mcmf_arfa}. In each Dinkelbach iteration, we solve the analog beamforming and hardware resource allocation problem for a fixed $q$.

\begin{algorithm}[htbp]
	\caption{Dinkelbach's algorithm with SV-based beamforming and MCMF-based ARFA.}
	\label{alg:dinkelbach_sv_mcmf_arfa}
	\begin{algorithmic}[1] 
		\STATE \textbf{Initialize:} Iteration $m=0$, convergence threshold $\xi > 0$, and initial EE $q^{(0)} > 0$.
		\REPEAT
		\STATE Solve problem \eqref{eq_prob2} for $q = q^{(m)}$ to obtain the analog combining weights $\{w_{\text{RF},l,i,n}^{(m)}\}$ via SV-based beamforming and the hardware configuration $\{a_l^{(m)}, n_l^{(m)}\}$ via MCMF-based ARFA.
		\STATE Calculate the current objective value: $F(q^{(m)}) = B R_{\text{sum}}^{(m)} - q^{(m)} P_{\text{total}}^{(m)}$.
		\STATE Update the EE parameter: $q^{(m+1)} = \frac{B R_{\text{sum}}^{(m)}}{P_{\text{total}}^{(m)}}$.
		\STATE Set $m = m + 1$.
		\UNTIL $\left|F(q^{(m)})\right| < \xi$.
	\end{algorithmic}
\end{algorithm}

\subsubsection{Dinkelbach's Algorithm}
The optimal energy efficiency $q^\star$ is achieved if and only if $\max F(q^\star) = 0$. This transformation allows us to iteratively update $q$ until convergence. However, solving the transformed problem \eqref{eq_prob2} remains challenging. Fortunately, we observe that the analog combining weights $\{w_{\text{RF},l,i,n}\}$ only affect the achievable rate $R_\text{sum}$ but not the power consumption $P_{\text{total}}$. Thus, for any given activation state $\{a_l, n_l\}$, optimizing the analog combiner $\{w_{\text{RF},l,i,n}\}$ can be naturally decoupled from the discrete hardware allocation.

\subsubsection{SV-Based Beamforming}
Under the D-HBF architecture, each active AP constructs the local analog beamforming matrix $\boldsymbol{W}_{\text{RF},l}^\star$ independently using its own channel estimate $\hat{\boldsymbol{H}}_l$. Specifically, for a given number of active RF chains $n_l$, $\boldsymbol{W}_{\text{RF},l}^\star$ is obtained by extracting the phases of $n_l$ leading singular vectors of $\hat{\boldsymbol{H}}_l$, as detailed in Section~\ref{subsec:achievable_rate}. It is worth noting that the analog combiner $\boldsymbol{W}_{\text{RF},l}^\star$ is solely determined by $\hat{\boldsymbol{H}}_l$ and $n_l$, independent of the system power $P_\text{total}$ and the Dinkelbach variable $q$. Consequently, the continuous phase variables $\{w_{\text{RF},l,i,n}\}$ are decoupled from the discrete optimization process, enabling pre-evaluation of the effective channel $\boldsymbol{H}_{\text{eff},l}$ and the corresponding marginal rate gains for the subsequent MCMF graph construction.

\subsubsection{MCMF-Based ARFA}
\begin{figure}[htbp]
	\centering
	\includegraphics[width=3.5in]{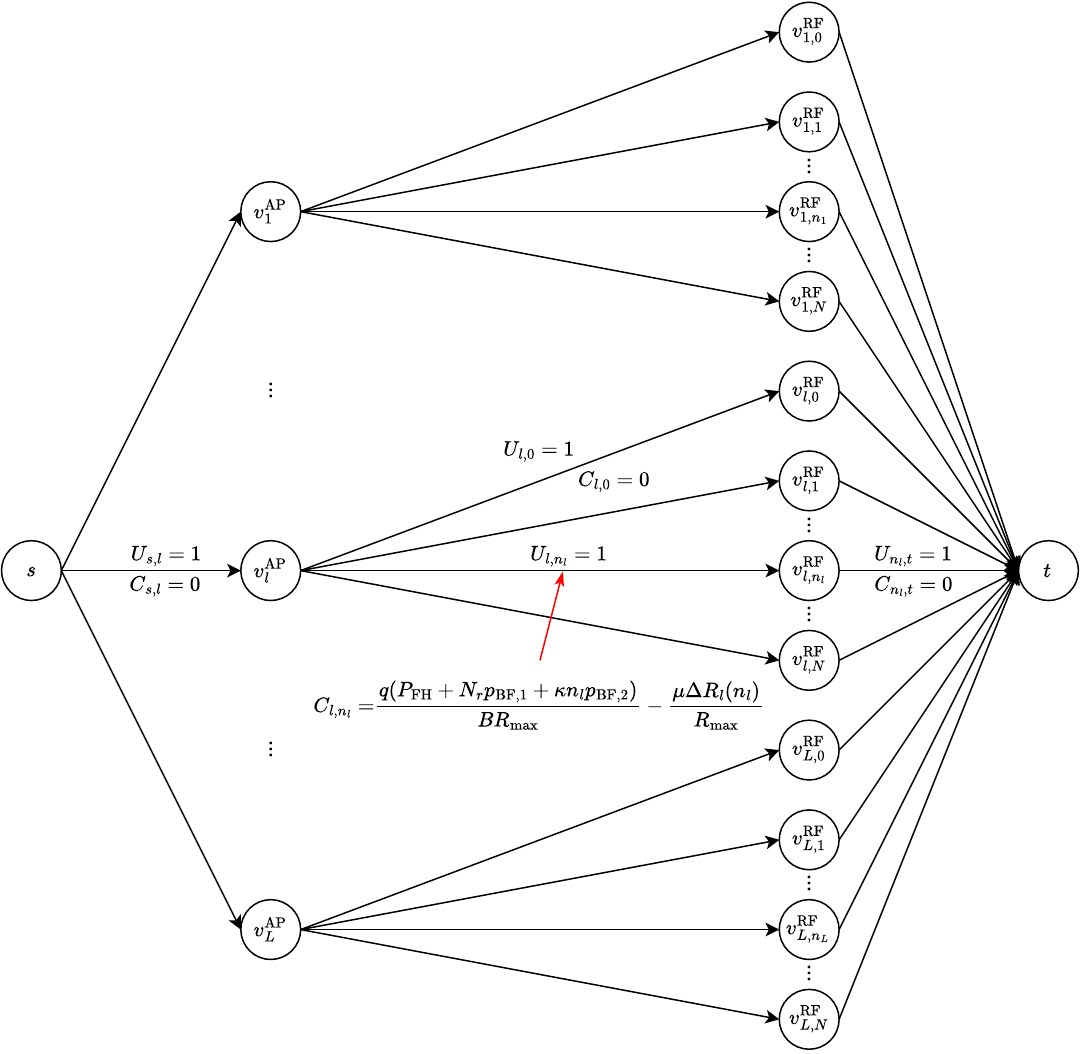}
	\caption{Directed graph for the MCMF-based resource assignment problem.}
	\label{mcmf_graph}
\end{figure}

With the optimal analog combiners $\{\boldsymbol{W}_{\text{RF},l}^\star\}$ fixed, the remaining subproblem reduces to finding the optimal hardware activation strategy $\{a_l, n_l\}$. As depicted in Fig.~\ref{mcmf_graph}, the resource assignment problem can be modeled as an MCMF problem over a digraph $G = (V, E)$, where $V$ and $E$ denote the sets of vertices (or nodes) and edges, respectively. For each directed edge $e=(u, v) \in E$, we define its capacity $U_{u,v}$ and generalized cost $C_{u,v}$. The edges of the constructed graph fall into three categories: 
\begin{itemize}
	\item AP selection edges: $(s, v_l^{\text{AP}})$ with $U_{s, l} = 1$ and $C_{s, l} = 0$.
	\item RF chain activation edges: $(v_l^{\text{AP}}, v_{l,n_l}^{\text{RF}})$ with $U_{l, n_l} = 1$ and $C_{l,n_l} = \mathds{1}(n_l \geq 1) [\frac{q(P_{\text{FH}} + N_r p_{\text{BF},1} + \kappa n_l p_{\text{BF},2})}{B R_\text{max}} - \frac{\mu\Delta R_l(n_l)}{R_\text{max}}]$, where $\kappa$ and $\mu$ denote the weighting factors of power costs and rate gains, respectively. Here, the rate increment $\Delta R_l(n_l)$ obtained by activating $n_l$ RF chains is approximated by $\Delta R_l(n_l) \approx \sum_{n=1}^{n_l} \log_2 (1 + \frac{\rho s_{l,n}^2}{\sigma_n^2})$, with $s_{l,n}$ representing the $n$-th dominant singular value of $\hat{\boldsymbol{H}}_l$, and $R_\text{max}$ serves as a normalization factor denoting the maximum sum SE achieved in full activation.
	\item Sink edges: $(v_{l,n_l}^{\text{RF}}, t)$ with $U_{n_l, t} = 1$ and $C_{n_l, t} = 0$.
\end{itemize}
The decision variables $\{a_l, n_l\}$ are uniquely determined by the optimal flow $f_{u,v}^\star$: if the unit flow traverses the edge $(v_l^{\text{AP}}, v_{l,n_l}^{\text{RF}})$ (i.e., $f_{l,n_l}^\star = 1$), the $l$-th AP is configured with $n_l$ active RF chains, where $n_l = 0$ implies the AP is inactive ($a_l = 0$) and $n_l \ge 1$ indicates $a_l = 1$. Since the constructed graph is a directed acyclic graph (DAG) with $\mathcal{O}(LN)$ nodes and edges, the formulated network flow problem can be efficiently solved using the network simplex algorithm.

\subsection{Complexity Analysis}
\label{subsec:cplx_anal}
Table~\ref{tab:complexity_comparison} compares the computational complexities of the aforementioned RF chain activation schemes. For a fair comparison, all schemes explicitly construct the MMSE digital combiners $\boldsymbol{W}_{\text{BB},l}^\star, \forall l$. Therefore, the complexities of the fixed-activation schemes with $n_l=N, \forall l$ and $n_l=\bar{n}, \forall l$ are $\mathcal{O}(L N K N_t N_r + 2 L K^2 N_t^2 N_r)$ and $\mathcal{O}(L \bar{n} K N_t N_r + 2 L K^2 N_t^2 N_r)$, respectively. To enable dynamic RF chain activation, the SV-based D-ARFA scheme incurs only a minor overhead for sorting singular values, and its dominant complexity remains $\mathcal{O}(L N K N_t N_r + 2 L K^2 N_t^2 N_r)$, where the first term accounts for the SVD computation of the $L$ channel matrices to obtain the analog combiners, while the second term corresponds to the digital combiner construction. In comparison, the complexity of the proposed MCMF-based ARFA scheme scales as $\mathcal{O}(L N K N_t N_r + 2 L K^2 N_t^2 N_r + I_\text{Dink} (L N)^3)$, where $I_\text{Dink}$ denotes the number of Dinkelbach iterations required for convergence, which is typically less than 10. Although the network simplex algorithm for the formulated MCMF problem introduces a higher polynomial complexity order, the graph size (i.e., $L + LN$ nodes) remains manageable for practical cell-free massive MIMO systems.
\begin{table}[ht]
	\centering
	\caption{Complexity Comparison}
	\label{tab:complexity_comparison}
	\resizebox{1.0\linewidth}{!}{
		\begin{tabular}{|m{1.2in}<{\centering}|m{2.3in}<{\centering}|}
			\hline
			{\textbf{Schemes}} & {\textbf{Complexities}} \\
			\hline
			Fixed-activation, $n_l=N$ & $\mathcal{O}(L N K N_t N_r + 2 L K^2 N_t^2 N_r)$ \\
			\hline
			Fixed-activation, $n_l=\bar{n}$ & $\mathcal{O}(L \bar{n} K N_t N_r + 2 L K^2 N_t^2 N_r)$ \\
			\hline
			SV-based D-ARFA & $\mathcal{O}(L N K N_t N_r + 2 L K^2 N_t^2 N_r)$ \\
			\hline
			MCMF-based ARFA & $\mathcal{O}(L N K N_t N_r + 2 L K^2 N_t^2 N_r + I_\text{Dink} (L N)^3)$ \\
			\hline
	\end{tabular}}
\end{table}

\section{Performance Evaluation}
\label{sec:perf_eval}
\subsection{Simulation Parameters}
\label{subsec:sim_param}
Simulations are carried out on a laptop equipped with an Intel Core i5-1235U CPU and 16 GB RAM to evaluate the performance of the proposed MCMF-based ARFA scheme. Specifically, we perform Monte Carlo simulations where APs and UEs are randomly distributed within a square area of $1000 \times 1000~\text{m}^2$. All APs and UEs are assumed to have identical antenna parameters, i.e., $G_r=15~\text{dBi}$, $G_t=0~\text{dBi}$, and $d_s=\lambda/2$. In the simulations, we consider a system scale of $L \in \{8,16,32,48,64\}$, $K=8$, $N_r=64$, $N_t=4$, $N=8$, and $\bar{n}=2$, where $\bar{n}$ is the average number of activated RF chains for the fixed-activation schemes. A practical mmWave system setting is adopted with carrier frequency $f_c=28~\text{GHz}$, subcarrier spacing $\mathrm{SCS}=120~\text{kHz}$, operating bandwidth $B=100~\text{MHz}$, and $\mathrm{NF}=9~\text{dB}$. Accordingly, the large-scale fading coefficients in \eqref{eq_pathloss} can be calculated using $n_\text{pl}=4.1$ and $\sigma_\text{sf}=7.6~\text{dB}$. Furthermore, the TDD parameters are set to $T_c=2~\text{ms}$ and $\tau_u=180~\text{symbols}$. 

With respect to the channel coefficients in \eqref{eq_channel}, we assume that the number of effective propagation paths between each AP and UE is identical, i.e., $P_{l,k}=3, \forall l,k$, reflecting the limited scattering nature of mmWave channels. The AoDs and AoAs are uniformly distributed in $[-\frac{\pi}{6}, \frac{\pi}{6}]$ and $[-\frac{\pi}{12}, \frac{\pi}{12}]$, respectively. The phases of the analog combiners are selected from the set $\Theta_\text{PS}=\{0, \frac{2\pi}{2^b}, \frac{4\pi}{2^b}, \dots, \frac{2(2^b-1)\pi}{2^b}\}$, where $b=4$ implies quantization bits of the PSs. To compute the total power consumption in \eqref{eq_power2}, the power parameters are assumed as follows: $\alpha=2~\text{bits}$, $C_\text{FH}=100~\text{Mbit/s}$, $P_\text{FH,max}=50~\text{W}$, $P_\text{fix}=0.825~\text{W}$, $p_\text{LNA}=20~\text{mW}$, $p_\text{ADC}=200~\text{mW}$, $p_\text{RF}=40~\text{mW}$, $p_\text{PS}=30~\text{mW}$, $p_\text{MIX}=0.3~\text{mW}$, $p_\text{SW}=5~\text{mW}$, $\eta = 0.3$, $P_\text{CP} = 1~\text{W}$, and $\rho=23~\text{dBm}$ ($\approx 0.2~\text{W}$). Furthermore, to account for imperfect CSI, the variance of the channel estimation error is set to $\varepsilon^2 = \epsilon \times \bar{g}$, where a moderate error factor $\epsilon=0.01$ is assumed, and $\bar{g} = \frac{G_tG_r}{KL} \sum_{l=1}^L \sum_{k=1}^K \frac{1}{\beta_{l,k}}$ represents the average large-scale fading coefficient~\cite{9780003}.

\subsection{Simulation Results}
\label{subsec:sim_res}
\begin{figure*}[htbp]
	\centering
	\subfloat[]{\includegraphics[height=1.33in]{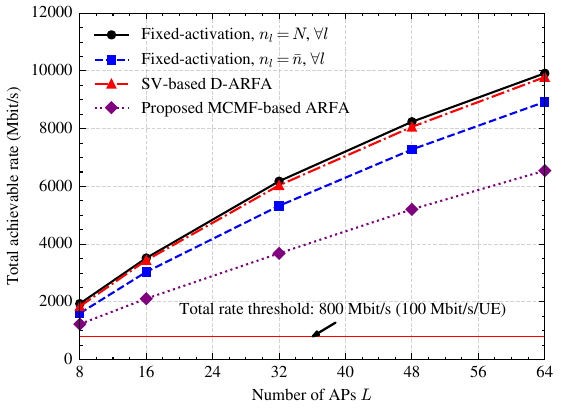}
		\label{total_achievable_rate}}
	\subfloat[]{\includegraphics[height=1.33in]{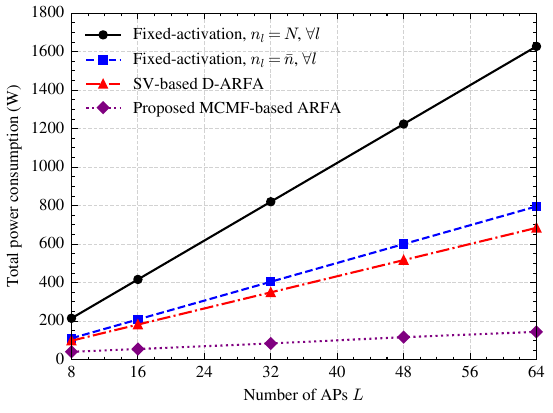}%
		\label{total_power_consumption}}
	\subfloat[]{\includegraphics[height=1.33in]{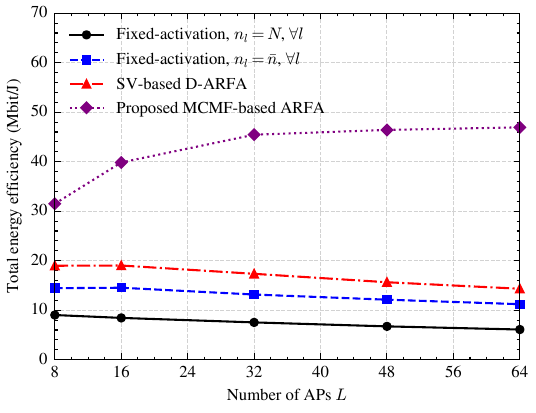}%
		\label{total_energy_efficiency}}
	\subfloat[]{\includegraphics[height=1.33in]{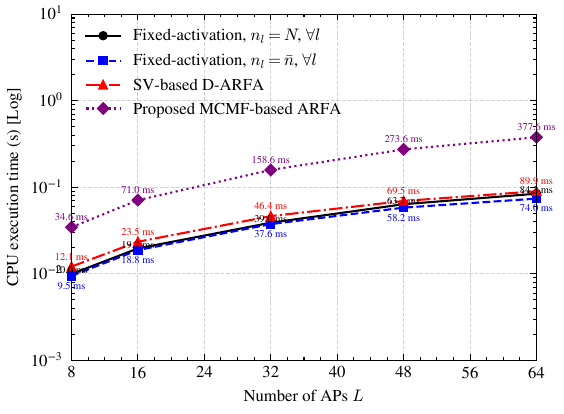}%
		\label{cpu_execution_time}}
	\caption{System performance comparison with simulation parameters $L \in \{8,16,32,48,64\}$, $K=8$, $N_t=4$, $N_r=64$, $N=8$, $\bar{n}=2$, and $\rho=23$ dBm. (a) Total achievable rate. (b) Total power consumption. (c) Total energy efficiency. (d) CPU execution time.}
	\label{performance_comparison}
\end{figure*}

Fig.~\ref{performance_comparison} compares the system performance, namely the achievable rate, power consumption, energy efficiency, and CPU runtime, versus the number of APs for various RF chain activation schemes with the simulation parameters above.

Numerical results in Fig.~\ref{performance_comparison}\subref{total_energy_efficiency} demonstrate that, notably for $L=48,64$, the proposed MCMF-based ARFA scheme yields remarkable EE improvements of roughly 600\%, 300\%, and 200\% over the fully-activated ($n_l=N$), partially-activated ($n_l=\bar{n}$), and SV-based D-ARFA schemes, respectively. In addition, the total EE of the proposed scheme improves with the number of APs, whereas existing benchmarks exhibit declining trends. Similarly in~Fig.~\ref{performance_comparison}\subref{total_power_consumption}, the total power consumption of our scheme is significantly lower than that of the other schemes, particularly as the number of APs increases. For instance, when $L=64$, the baseline schemes consume approximately 11.3, 5.5, and 4.7 times as much power as the proposed scheme, respectively.

Regarding the total achievable rate in Fig.~\ref{performance_comparison}\subref{total_achievable_rate}, the proposed approach is outperformed by existing benchmarks due to its aggressive power-saving strategy. Nevertheless, even in the worst-case scenario (i.e., $L=8$), the proposed scheme achieves a total rate of 1231.48~Mbit/s, with an average per-user rate of over 100~Mbit/s across the $K=8$ UEs, thereby ensuring satisfactory overall QoS. In situations with more stringent QoS requirements, the current formulation can be readily extended by incorporating a minimum per-user rate constraint $R_k \ge R_\text{min},\forall k$ into problem \eqref{eq_prob1} to further explore the EE-SE trade-off. Furthermore, the fronthaul constraint is explicitly omitted from the problem formulation since we assume error-free fronthaul links and $R_{\text{FH},l} \le C_{\text{FH},l},\forall l$ is always satisfied under the simulation parameters in Section~\ref{subsec:sim_param}. In addition, since $\tau_p \geq KN_t$ in low user-density environments such as our simulation setup, all UEs can utilize orthogonal pilot sequences to avoid inter-user interference. It remains essential to rigorously derive the channel estimation error by considering the pilot contamination effect in high user-density scenarios where $\tau_p < KN_t$ in future work.

Finally, the CPU execution time evaluated in Fig.~\ref{performance_comparison}\subref{cpu_execution_time} is consistent with the complexity comparison in Section~\ref{subsec:cplx_anal}. The runtime of the proposed algorithm is in the sub-second range, demonstrating its potential for real-time operation in realistic cell-free massive MIMO networks.

\section{Conclusion}
\label{sec:conclusion}
In this paper, we have investigated the energy-efficient design of next-generation wireless networks, with a particular focus on joint AP and RF chain activation in uplink mmWave cell-free massive MIMO systems. To maximize the total EE, the resource allocation is formulated as a non-convex optimization problem. By leveraging Dinkelbach's method to convert the fractional programming problem into its subtractive form, we propose an effective MCMF-based ARFA scheme. Numerical results demonstrate that the proposed approach significantly outperforms conventional fixed-activation and existing ARFA schemes in terms of total EE and power consumption, while maintaining a satisfactory achievable rate. Moreover, the proposed algorithm exhibits a manageable computational complexity with a sub-second CPU runtime, making it a promising candidate for energy-efficient and green communications in sustainable cell-free massive MIMO systems.


\bibliographystyle{IEEEtran}
\bibliography{reference.bib}

\end{document}